\documentclass{article}

\usepackage{amsmath}
\usepackage[a4paper,left=2.5cm,right=2.5cm]{geometry}
\usepackage[utf8]{inputenc}
\usepackage{graphicx}
\usepackage{xcolor}
\usepackage[caption=false,font=footnotesize]{subfig}
\usepackage{booktabs}

\providecommand{\keywords}[1]{\textbf{Keywords:} #1}

\graphicspath{{./images/}}

\title{\textbf{Fuel rod classification from Passive Gamma Emission Tomography (PGET) of spent nuclear fuel assemblies}}

\author{Riina~Virta$^{1,3,}$\thanks{Corresponding author: Riina Virta, email: \texttt{riina.virta@helsinki.fi}.},
Rasmus~Backholm$^2$, Tatiana A.~Bubba$^2$, Tapio~Helin$^4$, Mikael~Moring$^3$, \\ 
Samuli~Siltanen$^2$, Peter~Dendooven$^1$ and Tapani~Honkamaa$^3$}

\date{\small{%
$^1$Helsinki Institute of Physics, University of Helsinki, Finland. \\
$^2$Department of Mathematics and Statistics of the University of Helsinki, Finland. \\
$^3$Radiation and Nuclear Safety Authority (STUK), Finland. \\
$^4$School of Engineering Science of LUT University, Lappeenranta, Finland.}}

\begin{document}

\maketitle

\begin{abstract}
Safeguarding the disposal of spent nuclear fuel in a geological repository needs an effective, efficient, reliable and robust non-destructive assay (NDA) system to ensure the integrity of the fuel prior to disposal. In the context of the Finnish geological repository, Passive Gamma Emission Tomography (PGET) will be a part of such an NDA system. We report here on the results of PGET measurements at the Finnish nuclear power plants during the years 2017-2020. The PGET prototype device developed by IAEA and partners was used during 2017-2019, whereas an updated device was used in 2020.
The PGET device contains two linear arrays of collimated CdZnTe (CZT) gamma ray detectors installed opposite each other inside a torus. Gamma activity profiles are recorded from all angles by rotating the detector arrays around the fuel assembly that has been inserted into the center of the torus. Image reconstruction from the resulting tomographic data is defined as a constrained minimization problem with the function being minimized containing a data fidelity term and regularization terms. The activity and attenuation maps, as well as detector sensitivity corrections, are the variables in the minimization process. The regularization terms ensure that prior information on the (possible) locations of fuel rods and their diameter are taken into account. Fuel rod classification, the main purpose of the PGET method, is based on the difference of the activity of a fuel rod from its immediate neighbors, taking into account its distance from the assembly center. The classification is carried out by a support vector machine.
We report on the results for ten different fuel types with burnups between 5.72 and 55.0 GWd/tU, cooling times between 1.87 and 34.6 years and initial enrichments between 1.9 and 4.4\%. For all fuel assemblies measured, missing fuel rods, partial fuel rods and water channels were correctly classified. Burnable absorber fuel rods were classified as fuel rods. On rare occasions, a fuel rod that is present was falsely classified as missing. The further development of the image reconstruction method is discussed.
We conclude that the combination of the PGET device and our image reconstruction method provides a reliable base for fuel rod classification. The method is thus well-suited for nuclear safeguards verification of fuel assemblies in Finland prior to geological disposal.
 \end{abstract} 


\section{Introduction}

Finland will start disposal of spent nuclear fuel in a deep geological repository around the mid-2020's, the first country in the world to do so. The construction of the underground facility is ongoing at Olkiluoto, Eurajoki. For safeguards purposes, the disposal needs an effective, efficient, reliable and robust non-destructive assay (NDA) system for spent nuclear fuel verification. The combination of Passive Gamma Emission Tomography (PGET) and Passive Neutrino Albedo Reactivity (PNAR)~\cite{Tobin2018} will be used for this. At the end of 2017, the International Atomic Energy Agency (IAEA) approved PGET for the verification of spent nuclear fuel. Other state-of-the-art methods (e.g. Fork detectors~\cite{Vaccaro2018}) can only detect a gross deviation of material in the fuel assembly, but PGET has been demonstrated to accomplish reliable rod-level detection~\cite{Honkamaa2014,Mayorov2018}. This is crucial for ensuring effective nuclear safeguards of the final repository.

First PGET images of fuel assemblies of different types and with varying cooling times and burnups were published in~\cite{White2018} and~\cite{Belanger-Champagne2019}. These results indicated the need for improved image reconstruction and analysis methods. Recently, we proposed a method in which attenuation and activity images are simultaneously reconstructed from the data by formulating the reconstruction as a constrained minimization problem and solving it with a Levenberg-Marquardt type of method~\cite{Backholm2020}. In the present study, we apply this method to PGET data taken at the two Finnish nuclear power plants (NPPs). We show that high-quality results are produced, enabling to detect a single missing rod in a wide range of different assembly types and parameters. The results are significant in both the context of Finnish safeguards and the global context of nuclear fuel disposal.

\section{Materials and methods}

\subsection{Imaged spent nuclear fuel}

PGET measurements were performed at the two Finnish NPPs, Olkiluoto (OL) and Loviisa (LO), during the years 2017-2020. A total of 78 individual assemblies and 7 non-fuel items were measured. The assembly types were VVER-440 at Loviisa and 9 BWR type assemblies at Olkiluoto (SVEA-64, SVEA-96, SVEA-96 OPTIMA, SVEA-100, ATRIUM10, GE12, GE14, 9x9-1AB and 8x8-1). The assemblies were chosen to cover a wide range of operating parameters: burnup from 5.72 to 55.0 GWd/tU, cooling time from 1.87 to 34.6 years and initial enrichment from 1.9 to 4.4 \%. 
We report results from a selection of  measurements from the campaigns in Olkiluoto in 2017 and 2019 (OL17, OL19) and Loviisa in 2018 and 2020 (LO18, LO20). The main fuel characteristics are presented in Table~\ref{tab:assembly_characteristics}. The measured assemblies were chosen to best reflect the strengths and future development areas of our software and to represent a wide range of fuel characteristics and specifics (such as 1 to 3 completely removed fuel rods, burnable absorber rods or partial rods), measurement campaigns and assembly types.

\subsection{PGET device}
The PGET device (see~\cite{White2018} for details) contains 174 highly collimated CdZnTe (CZT) gamma-ray detectors arranged in 2 linear arrays on opposite sides of a torus. These arrays are rotated around the fuel assembly to collect data from all angles. All measurements were done underwater in spent fuel ponds. Measurements were conducted with different numbers of projection angles and measurement times per angle. Some assemblies were measured at different vertical, horizontal and rotational positions. In this work, we show results from measurements with 360 angles and 800 ms projection time per angle (OL17, OL19, LO18) or 924 ms projection time per angle (LO20). Data were collected in four energy windows. The lowest two windows were 400-600 keV and 600-700 keV. The third and fourth windows were 700-1500 keV and above 1500 keV for OL17, OL19 and LO18, and 700-2000 keV and 2000-3000 keV for LO20. The choice of these windows is related to the gamma peaks of the radioactive nuclei present. All image reconstructions shown in this work are from the 600-700 keV window which contains the 661 keV gamma peak from $^{137}$Cs, the most abundant gamma ray emitter in spent fuel.

The so-called prototype PGET device was used for the campaigns in 2017, 2018 and 2019. However, some individual detectors were replaced in-between campaigns. The campaign at Loviisa in 2020 used a new PGET device with a more compact design for easier handling and a slightly optimized collimator.

\begin{table*}[t]
\centering
\label{tab:assembly_characteristics}
\begin{tabular}{@{}ll@{}r@{\;\;\;}r@{\;\;\;}ll@{}}
\toprule
\# & Type           & BU {[}GWd/tU{]} & CT {[}a{]} & Campaign & Characteristics   \\ \midrule
1  & VVER-440       & 55.0                               & 4.6                            & LO20     & 3 missing (corner), burnable absorber rods   \\
2  & VVER-440       & 42.0                                & 5.6                            & LO20     & 1 missing (corner)                           \\
3  & VVER-440       & 43.0                                & 2.7                            & LO20     & Burnable absorber rods                       \\
4  & VVER-440       & 22.8                                & 27.6                           & LO18     & Activity and attenuation gradient            \\
5  & SVEA-64        & 32.6                                & 20.7                           & OL19     & 2 rods in the reactor for 2/4 fuel cycles    \\
6  & SVEA-64        & 32.9                                & 20.7                           & OL19     & Intra-rod activity differences               \\
7  & SVEA 96        & 40.7                                & 8.9                            & OL17     & 2 missing  (low-left and low-right quarters) \\
8  & SVEA-96 Optima & 39.8                                & 13.7                           & OL19     & Measurements at two heights                  \\
9  & ATRIUM10       & 49.7                                & 7.9                            & OL19     & 2 missing, measurements at two heights       \\
10 & 8x8-1          & 18.6                                & 34.6                           & OL19     & Long CT                                      \\
11 & 9x9-1AB        & 35.0                                & 20.7                           & OL19     & Different assembly type                      \\
12 & GE12           & 43.1                                & 11.7                           & OL19     & Measurements at two heights                  \\ \bottomrule
\end{tabular}
\caption{Measured fuel assemblies and their characteristics: assembly type, burnup, cooling time and measurement campaign (at Loviisa (LO) and Olkiluoto (OL) during the years 2017-2020). The characteristics are from the licence-holder's declaration.}
\end{table*}

\subsection{Data analysis and image reconstruction}
\label{sec:analysis_reconstruction}

\subsubsection{Image reconstruction algorithm}    
The core idea of our image reconstruction strategy is to recover simultaneously the activity and the attenuation map from the data. To enhance performance with real data, we also tweak the detector sensitivity correction during the reconstruction process. This is all attained by formulating the reconstruction task as a constrained minimization problem where the activity and the attenuation maps, as well as the sensitivity correction coefficients, are the variables, and the function being minimized consists of a least squares data fit term and regularization terms~\cite{Engl1996}. Namely, the minimization problem takes the following form:  
\begin{equation}
\min_{\lambda,\mu,c} \; \Big\{
    \|H(\mu) \lambda - C(c)s\|_2^2 + \alpha_{\lambda} \|R_\lambda \lambda \|_2^2 + 
    \alpha_{\mu} \|R_\mu \mu \|_2^2  
    + \alpha_{c} \|\log(c)\|_2^2 +
    \alpha_{s} \| \mathbf{1}^T (s- C(c) s) \|_2^2 
    \Big\}
\label{eq:MinProbl}     
\end{equation}
with the bounds
\begin{equation}
A \begin{bmatrix} \lambda \\ \mu \end{bmatrix}
    \leq b.
\label{eq:Bounds}     
\end{equation}

Here $\lambda$ and $\mu$ are the vector forms of the discrete activity and attenuation maps, respectively, and the vector $c$ consists of the coefficients used to correct for the detector sensitivity differences in the data sinogram $s$. The least squares term $\| H(\mu)\lambda - C(c)s\|_2^2$ measures how well $\lambda$, $\mu$ and $c$ fit the data $s$. The rest of the terms are regularization terms, which can be understood to incorporate some kind of a priori information in the reconstruction process, namely, they predispose the algorithm towards certain kinds of solutions. The regularization parameters $\alpha_\lambda$, $\alpha_\mu$, $\alpha_c$ and $\alpha_s$ balance the contribution of each term.

In more detail, in the data fit term ${\| H(\mu)\lambda - C(c)s\|_2^2}$, $H(\mu)$ is the system matrix depending on the attenuation map $\mu$: a detailed description of how to implement the system matrix $H(\mu)$ can be found in~\cite{Backholm2020}. The product $H(\mu)\lambda$ is the forward projection, i.e., a sinogram simulated using $\lambda$ and $\mu$. The system matrix models the effects of attenuation and collimator blurring. The spatial responses are computed based on the given dimensions of the device and assuming an opaque collimator. $C(c)$ is a diagonal matrix formed from the detector sensitivity correction coefficients $c$ so that in the product vector $C(c)s$, which is the sensitivity-corrected sinogram in vector form, all the elements of $s$ corresponding to values from one detector are multiplied by one coefficient in $c$. 
 
The effect of the regularization terms $\alpha_{\lambda} \|R_\lambda \lambda \|_2^2$ and $\alpha_{\mu} \|R_\mu \mu \|_2^2$ depends on the choice of the matrices $R_\lambda$ and $R_\mu$. From the two types of regularization terms introduced in~\cite{Backholm2020}, to which we refer for the precise mathematical formulation, we considered in this paper only the geometry aware prior. We consider this prior to be suitable in the context of verification of spent nuclear fuel. This choice for the matrices $R_\lambda$ and $R_\mu$ assumes that the locations and diameters of possible rods, whether they are actually present or not, are known. In practice, this prior asserts that the solution should look approximately like it is made out of rods, each having a uniform activity, with the predefined diameters in the predefined locations. A rod can be missing in these solutions by having zero activity and the attenuation of water.

The term $\alpha_{c} \|\log(c)\|_2^2$ penalizes large absolute values of $\log(c)$ and so prefers solutions where the coefficients in $c$ are close to one, i.e., the corrections made are not large. The last term ${\alpha_{s} \| \mathbf{1}^T (s- C(c) s) \|_2^2}$, where $\mathbf{1}$ is a vector of ones, requires that the sum of all counts in the corrected sinogram $C(c)s$ is close to the sum of all the counts in the sinogram $s$. The scope of this term is to keep the same ``overall scale'' of the sinogram after correction. The data sinogram $s$ has actually already undergone a preliminary detector sensitivity correction, and the role of the coefficients $c$ is only to fine-tune this. The approach we first introduced in~\cite{Backholm2020} did not include the correction coefficients $c$. 

\begin{figure}[!tb]
    \centering
    \begin{tabular}{c@{\;}c}
    \rotatebox{90}{\hspace{2.5em} Activity $\lambda$ (relative)}
    & \includegraphics[scale=0.25]{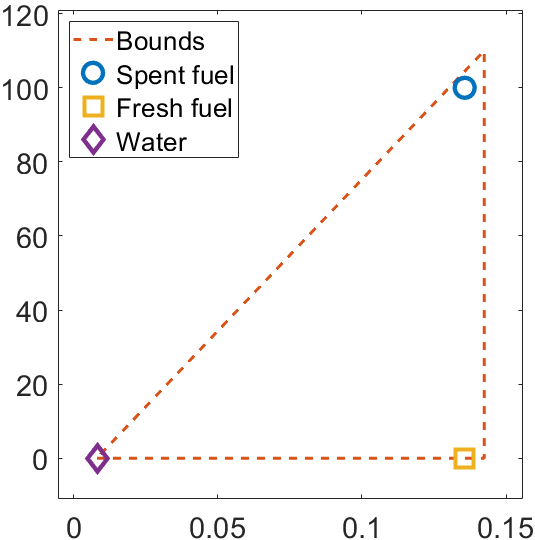} \\
    & \hspace{1.0em} Attenuation coefficient $\mu$ (1/mm)
    \end{tabular}
    \caption{Example of the bounds for the activity values and attenuation coefficients (for 661 keV gamma rays) used in the minimization process. The values inside the triangle are allowed.}
    \label{fig:Bounds}
\end{figure}

The bounds~\eqref{eq:Bounds} on the activity values and attenuation coefficients used in the minimization process are such that they exclude the possibility of a material with high activity and low attenuation coefficient, which is a physically unlikely case. In practice, to define these bounds, one must give lower and upper bounds for the attenuation coefficients and an upper bound for activity values (the lower bound for activity is always considered to be zero).
The bounds can be better understood by visualizing them in the attenuation-activity plane, as depicted in Fig.~\ref{fig:Bounds}, where the allowed values form a triangle.

To estimate the activity (and attenuation) bounds and to build the matrices  $R_\lambda$ and $R_\mu$ for the geometry aware prior, rod locations and diameters are needed. We retrieve this information by identifying the assembly type from an ``off-the-shelf'' filtered back-projection (FBP) reconstruction, and then using the known grid and rod dimensions for that type of assembly. In case the assembly type has water channels, these are not assumed to be known: rather a full grid of rods is assumed.

In practice, the bounds for attenuation coefficients are estimated by considering the measurement energy window and the materials assumed to be imaged. Once the attenuation coefficients for water and rods are estimated, and locations and diameters for the rods are computed from having identified the assembly type, 
the upper bound for activity values is estimated by simulating a sinogram using rods with some uniform activity value $a$. The upper bound for activity values is then set so that the ratio of $a$ and the maximum value of the simulated sinogram is the same as the ratio of the upper bound and the maximum value of the data sinogram.

Finally, the minimization problem~\eqref{eq:MinProbl} is solved iteratively using a Levenberg-Marquardt type of algorithm~\cite{Kaltenbacher2008}, that takes the bounds~\eqref{eq:Bounds} into account~\cite{Kelley1999}. All the reconstructions are computed using 120 evenly spaced measurement angles (every third angle from the total of 360 measurement angles), which provides a good balance between computational efficiency and quality of the reconstruction in order to perform a reliable rod classification.

\subsubsection{Rod classification}
To classify rods into missing or present ones, the basic idea is to consider the difference of a rod's activity from the average activity of its immediate neighbors plotted against its distance from the assembly center. Rod activity values are computed as a weighted average of the values of all the pixels that consist at least partly of the rod. The weights are proportional to the fraction of the area of the pixel that is covered by the rod, namely, the border pixels contribute less to the average. The sum of the weights is not normalized to one and as the classification is done on rod-level, the values for individual rods can exceed the bounds, as can be seen in Fig.~\ref{fig:12_bounds_neighbors}.

The classification is carried out by training a support vector machine~\cite{cristianini2000} on the reconstructions of training data sets from varied assemblies, including mock-up fuel and real spent fuel assemblies with different rod placements and intensities. Data from assemblies of mock-up fuel constructed from neutron-activated cobalt rods containing $^{60}$Co were measured at the Atominstitut at the Technical University of Vienna where the PGET system is prepared for spent-fuel measurements~\cite{White2018}. At first, rather than classifying all the rods at once, we begin classifying them as missing one by one starting from the most likely case. After classifying a rod as missing, the differences the classification is based on are recalculated by discarding from the calculations of the neighbor average all the rods classified as missing at that point. This prevents the missing rods from bringing down the neighbor average. 

When a rod shows to have at least a couple of missing neighbors, we compute another neighbor average of only the rods that are classified as missing and use that as well: if a rod's activity is close to the average of its missing neighbors' activities, it is classified as missing. This comparison allows classification of larger missing rod areas where rods might not have neighboring rods present. For the second comparison, a second support vector machine was trained using the same reconstructions as training data as was used for the first support vector machine. 

The final result is a plot where missing and present rods are separated rather than by a straight line, by a possibly more complex curve.

\section{Results}
\label{sec:results}

% 3 (2017)
% VVER-440 (BU 43.7 GWd/tU, CT 14.6 a)
% 3 missing rods
% 8 (2017)
% VVER-440 (BU 0.572 GWd/tU, CT 22.1 a)
% low BU

Good quality reconstructions are needed as a basis for detecting anomalies and accurately classifying spent fuel rods. In the following we present results from data from a selection of measurements (see Table~\ref{tab:assembly_characteristics}) and demonstrate the ability to detect missing rods, burnable absorber rods, water channels and intra-rod activity differences.

\subsection{Missing and abnormal rod detection}
\label{sec:missing}

\begin{figure}[h!]
\centering
\begin{tabular}{@{}c@{\quad}c@{\quad}c@{}}
\includegraphics[scale=0.35]{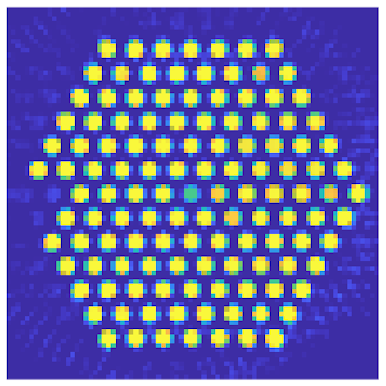}     
& \includegraphics[scale=0.35]{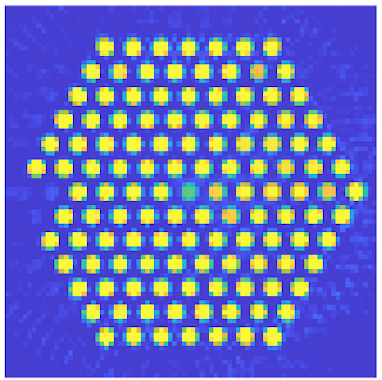}
& \includegraphics[scale=0.35]{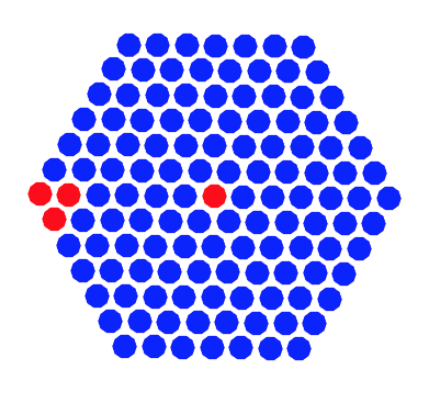} \\
\includegraphics[scale=0.35]{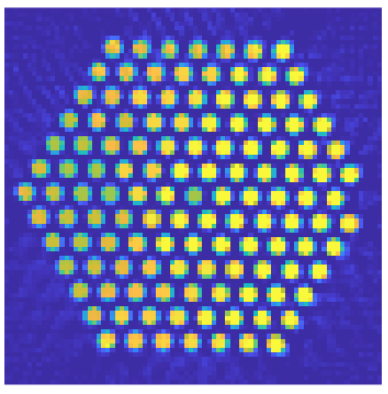}     
& \includegraphics[scale=0.35]{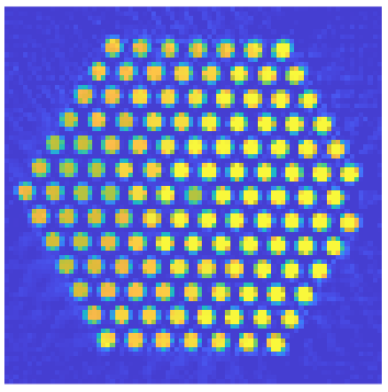}   
& \includegraphics[scale=0.35]{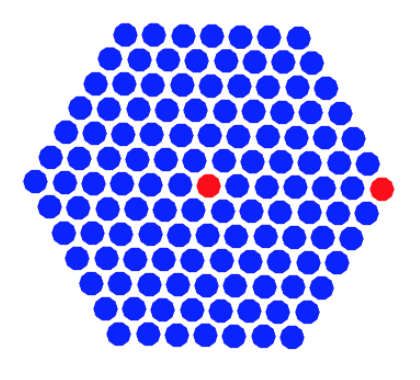}
\end{tabular}
     \caption{Activity (left column) and attenuation (middle column) reconstructions and classification into missing (red) and present (blue) rods (right column) for two VVER-440 assemblies. Assembly \#1 in the top row has three missing rods, a central water channel and 5 burnable absorber rods near the corners. Assembly \#2 in the bottom row has one missing rod and a central water channel.} 
     \label{fig:12&21}
\end{figure}

Missing rod detection in hexagonal VVER-440 assemblies is demonstrated for two assemblies in Fig.~\ref{fig:12&21} with activity and attenuation reconstructions and a rod classification figure. Assembly \#1 in the top row (BU 55.0 GWd/tU, CT 4.6 a) has three missing fuel rods and five burnable absorber rods (see Section~\ref{sec:burnable_absorber}) and assembly \#2 in the bottom row (BU 42.0 GWd/tU, CT 5.6 a) has one missing rod. In both cases the missing rods are clearly visible in the reconstructions and correctly classified by the algorithm. The central water channel is also classified as missing in both cases (see Section~\ref{sec:water_channel}).

% 12 (#1)
% VVER-440 (BU 55.0 GWd/tU, CT 4.6 a)
% 3 missing rods

% 21 (#2)
% VVER-440 (BU 42.0 GWd/tU, CT 5.6 a)
% 1 missing rod

\begin{figure}[h!]
\centering
\begin{tabular}{@{}c@{\quad}c@{\quad}c@{}}
\includegraphics[scale=0.5]{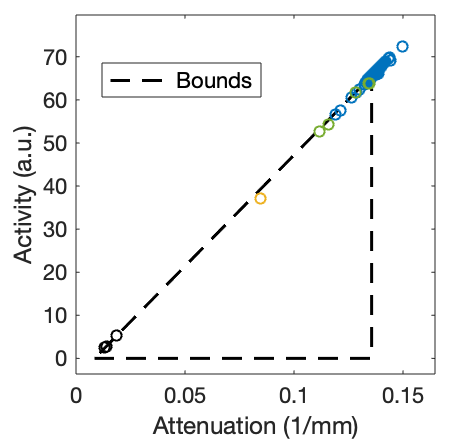}     
& \includegraphics[scale=0.5]{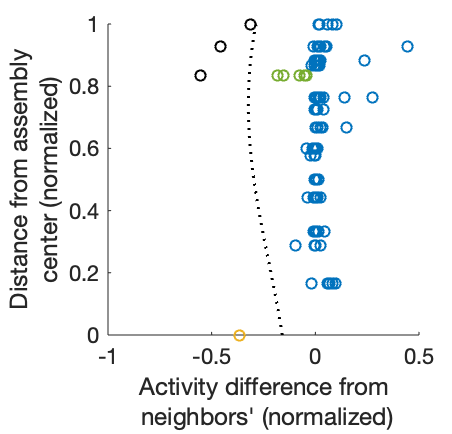}    
\end{tabular}
     \caption{Rod classification plots for the VVER-440 assembly \#1 (see also Fig.~\ref{fig:12&21}, top row). Linear bounds and average rod values are shown in the attenuation-activity plane on the left and rod activity difference from the neighbors as a function of the  distance from the assembly center is shown on the right. Circles represent individual rods and colors denote the ground truth rod type (blue for present, yellow for water channel, black for missing and green for burnable absorber rod). The dotted line on the right represents the classification border.}
     \label{fig:12_bounds_neighbors}
\end{figure}

Fig.~\ref{fig:12_bounds_neighbors} shows two classification metric plots for the VVER-440 assembly \#1 with three missing rods and five burnable absorber rods (see also Fig.~\ref{fig:12&21}, top row). Each circle represents a rod position and the color denotes the ground truth rod type. The rods get grouped by their characteristics and the rod classification is based on the kind of plots on the right. On the left, the three missing rods show low attenuation and low activity, and on the right they have a negative activity difference compared to their neighboring rods. The burnable absorber rods get grouped with the present rods but the water channel position deviates from the rest of the rods and is correctly classified as missing. 

% 12, 3 missing, poison rods

\begin{figure}[h!]
\centering
\begin{tabular}{@{}c@{\quad}c@{\quad}c@{}}
\includegraphics[scale=0.285]{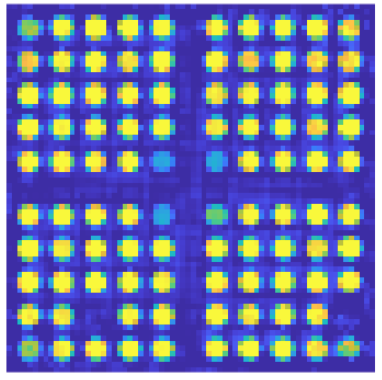}     
& \includegraphics[scale=0.285]{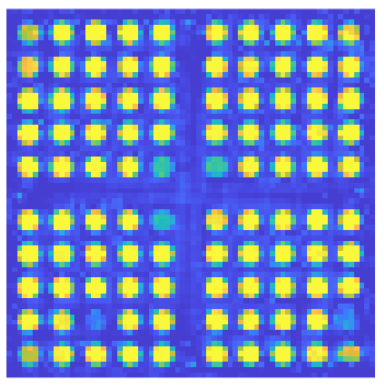}
& \includegraphics[angle=-0.3,scale=0.285]{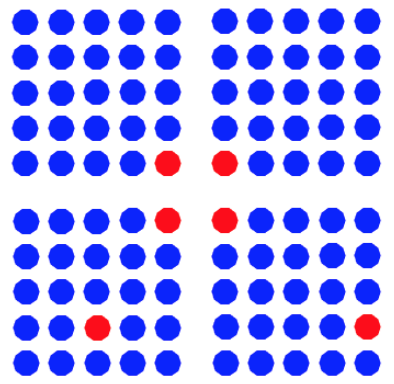} \\
\includegraphics[scale=0.35]{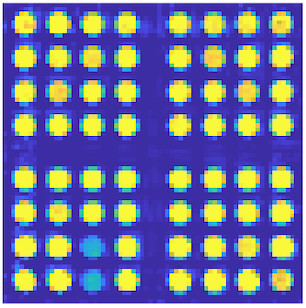}     
& \includegraphics[scale=0.35]{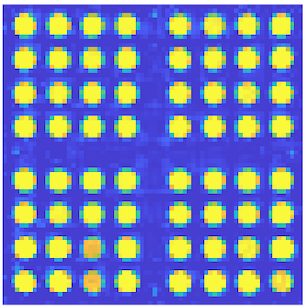}
& \includegraphics[scale=0.34]{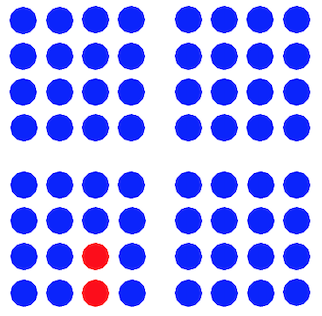}
\end{tabular}
     \caption{Activity (left column) and attenuation (middle column) reconstructions and classification into missing (red) and present (blue) rods (right column). The top row shows the SVEA-96 assembly \#7 with two missing rods and a water channel (four center-most rods) and the bottom row shows the SVEA-64 assembly \#5 with two rods that have been in the reactor for only 2 out of the normal 4 fuel cycles.} 
     \label{fig:O1&O28}
\end{figure}

Activity and attenuation reconstructions and rod classification for the SVEA-96 assembly \#7 (BU 40.7 GWd/tU, CT 8.9 a) with two missing rods is shown in Fig.~\ref{fig:O1&O28}, top row. The four innermost rods are a part of the water channel. The bottom row shows the reconstructions and the classification for the SVEA-64 assembly \#5 (BU 32.6 GWd/tU, CT 20.7 a) with two fuel rods that have been in the reactor for only two out of the normal four fuel cycles and thus have a different burnup than the other rods in the assembly.

% O1 (#7)
% SVEA-96 (BU 40.7 GWd/tU, CT 8.9 a)
% 2 missing rods
% O28 (#5)
% SVEA-64 (BU 32.6 GWd/tU, CT 20.7 a)
% 2 rods in the reactor 2/4 cycles

\begin{figure}[h!]
\centering
\begin{tabular}{@{}c@{\quad}c@{\quad}c@{}}
\includegraphics[scale=0.35]{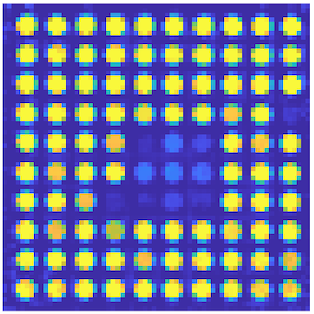}
&  \includegraphics[scale=0.35]{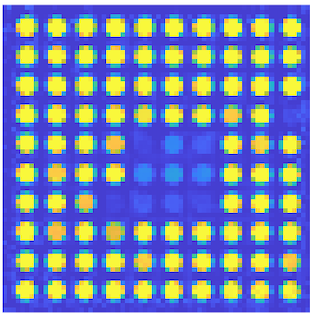}
& \includegraphics[scale=0.35]{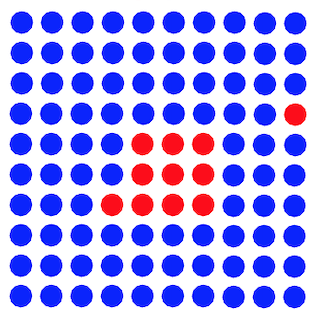} \\
\includegraphics[scale=0.35]{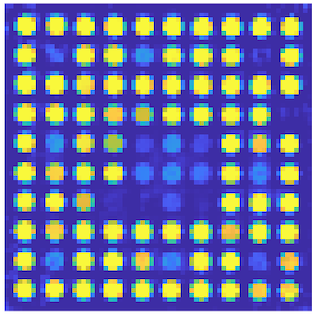} 
&  \includegraphics[scale=0.35]{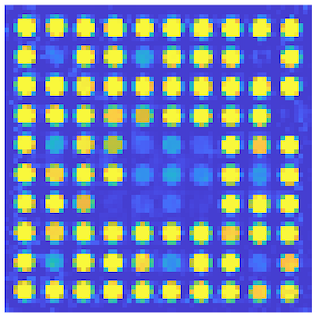}
& \includegraphics[scale=0.35]{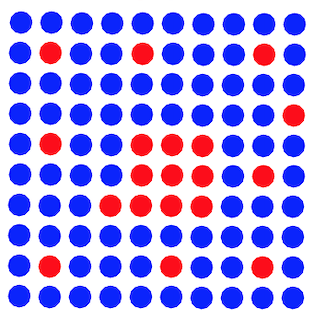}
\end{tabular}
\caption{Activity (left column) and attenuation (middle column) reconstructions and classification into missing (red) and present (blue) rods (right column) for the ATRIUM10 assembly \#9 with two missing rods. The top row reconstructions are from data measured at normal height and the bottom row at the upper position, where the partial fuel rods disappear from view.}
\label{fig:O23}
\end{figure}

An ATRIUM10 assembly has a 3x3 water channel and eight partial rods, which start at the bottom of the assembly and have a length of 2/3 compared to the rest of the rods. Fig.~\ref{fig:O23} shows attenuation and activity reconstructions and rod classification for the ATRIUM10 assembly \#9 (BU 49.7 GWd/tU, CT 7.9 a) at both the normal measurement height as well as at the upper position 1.5 meters higher where the partial rods disappear from view. In the data collected at the higher position, the partial rod positions are classified as missing. This assembly also has two missing rods which are correctly classified at both measurement heights.

% O23 (#9)
% ATRIUM10 (BU 49.7 GWd/tU, CT 7.9 a)
% 1 missing rod

A total of eight assemblies with one to three missing rods were measured and in all cases the missing rods were accurately classified by the algorithm. Here we have shown results from five of these assemblies.

\subsection{Burnable absorber detection}
\label{sec:burnable_absorber}

Some fuel assemblies contain fuel rods with added burnable absorber (usually Gd) used to balance the reactivity of the reactor during operation. Once the fuel assembly has reached the end of its operational lifetime, in an optimal case the burnable absorbers have reached the relative burnup of the other rods in the assembly. However, if the assembly has been removed before the absorber material has reached this level, the burnable absorber rods will show up as less active in the reconstructions.

\begin{figure}[h!]
\centering
\begin{tabular}{@{}c@{\quad}c@{\quad}c@{}}
\includegraphics[scale=0.35]{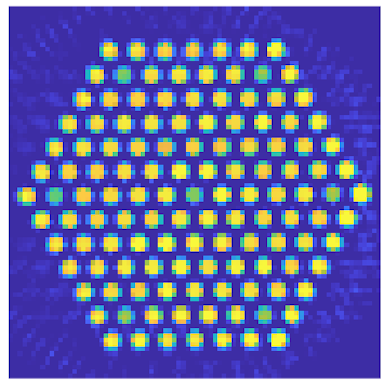}     
& \includegraphics[scale=0.35]{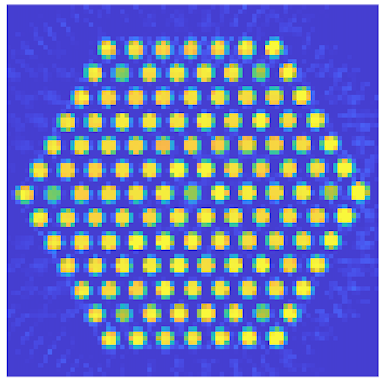}
& \includegraphics[scale=0.35]{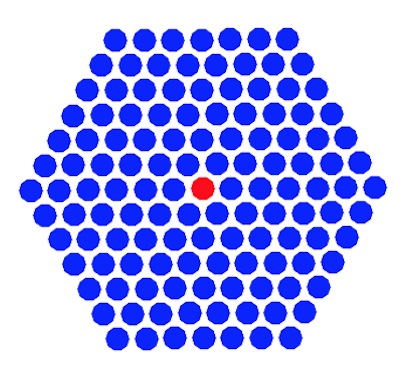}
\end{tabular}
     \caption{Activity (left) and attenuation (middle) reconstructions and classification into missing (red) and present (blue) rods (right) for the VVER-440 assembly \#3 with a central water channel and 6 burnable absorber rods near the corners. The water channel is classified as missing but the burnable absorber rods are not.}
     \label{fig:20}
\end{figure}

To ensure accurate detection of missing fuel rods and to avoid false alarms, the burnable absorber rods should not be classified as missing by the classification algorithm. Our results show correct classification of burnable absorber rods as present, as demonstrated in Fig.~\ref{fig:12&21} (upper row), where the VVER-440 assembly \#1 with five burnable absorber rods is shown. Fig.~\ref{fig:20} shows the VVER-440 assembly \#3 (BU 43.0 GWd/tU, CT 2.7 a) and similarly, the burnable absorber rods are somewhat visible in the reconstructions but not classified as missing.

% 20 (#3)
% VVER-440 (BU 43.0 GWd/tU, CT 2.7 a)
% burnable absorber rods, center water channel noticed

A total of eight VVER-440 assemblies with burnable absorber rods were measured and all burnable absorber rods were accurately classified as present by the algorithm. Here  we  have shown  results  from  two  of  these assemblies.

\subsection{Water channel and partial rod detection}
\label{sec:water_channel}

The accuracy of the algorithm in detecting missing fuel rods was also tested by its ability to correctly classify water channels and partial rod positions as missing. This is demonstrated with a variety of different assembly types. For VVER-440 type assemblies, the center water channel is accurately classified as missing in Fig.~\ref{fig:12&21} and Fig. ~\ref{fig:20} (assemblies \#1, \#2 and \#3). For a SVEA-96 assembly, the water channel classification can be seen in Fig.~\ref{fig:O1&O28} (top row, assembly \#7) and for an ATRIUM10 in Fig.~\ref{fig:O23} (assembly \#9).

\begin{figure}[h!]
\centering
\begin{tabular}{@{}c@{\quad}c@{\quad}c@{}}
\includegraphics[scale=0.35]{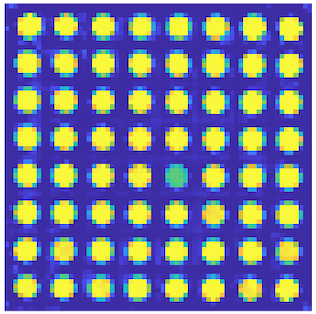}     
& \includegraphics[scale=0.35]{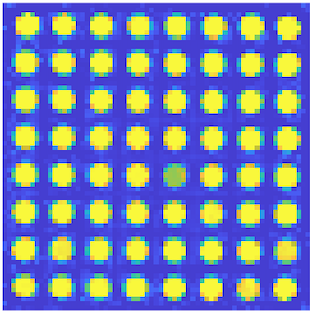}
& \includegraphics[scale=0.35]{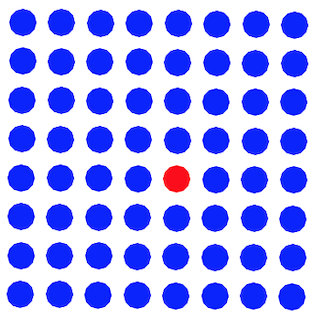}\\
\includegraphics[scale=0.35]{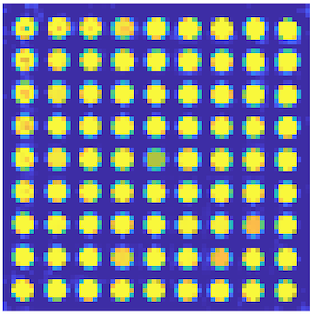}     
& \includegraphics[scale=0.35]{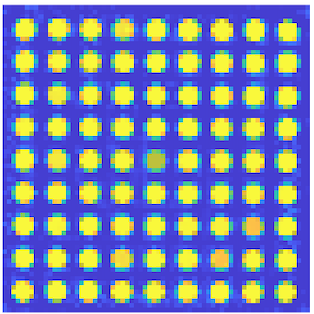}   
& \includegraphics[scale=0.35]{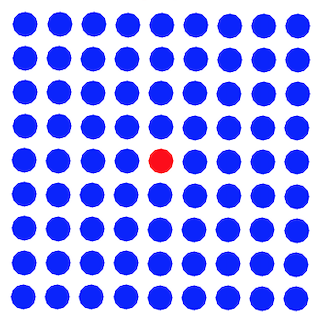}
\end{tabular}
     \caption{Activity (left column) and attenuation (middle column) reconstructions and a classification into missing (red) and present (blue) rods (right column) for the 8x8-1 assembly \#10 (top row) and the 9x9-1AB assembly \#11 (bottom row). Both reconstructions show accurate classification of the water channel near the center.}  
      \label{fig:O37&O41}
\end{figure}

Fig.~\ref{fig:O37&O41} shows the water channel classification for the 8x8-1 assembly \#10 (BU 18.6 GWd/tU, CT 34.6 a, top row) and the 9x9-1 assembly \#11 (BU 35.0 GWd/tU, CT 20.7 a, bottom row). In both assemblies, the water channel near the center is correctly classified as missing. The former reconstruction also demonstrates the ability to gain accurate results from a long-cooled fuel assembly, which is very relevant in the context of a deep geological repository. Assemblies with even longer cooling times will be expected once the disposal starts.

% O37 (#10)
% 8x8-1 (BU 18.6 GWd/tU, CT 34.6 a)
% high CT

% O41 (#11)
% 9x9-1AB (BU 35.0 GWd/tU, CT 20.7 a)

\begin{figure}[h!]
\centering
\begin{tabular}{@{}c@{\quad}c@{\quad}c@{}}
\includegraphics[scale=0.35]{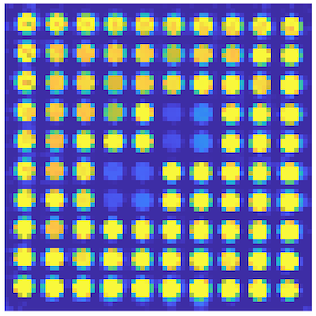}
&  \includegraphics[scale=0.35]{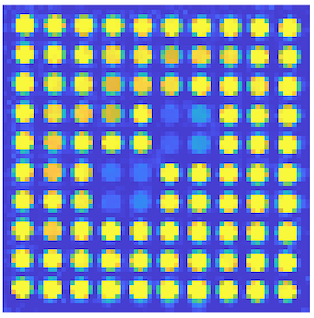}
& \includegraphics[scale=0.35]{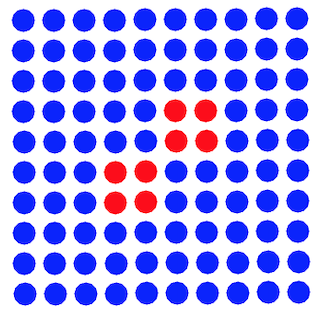}\\
\includegraphics[scale=0.35]{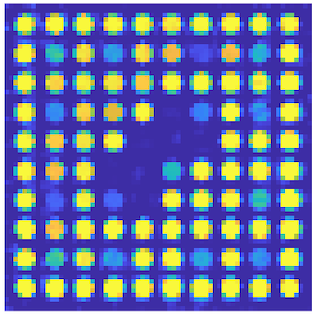} 
&  \includegraphics[scale=0.35]{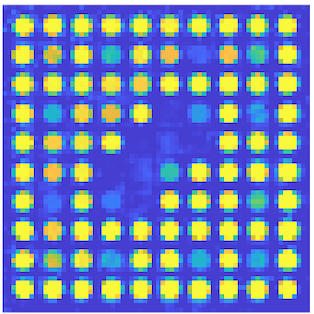}
& \includegraphics[scale=0.35]{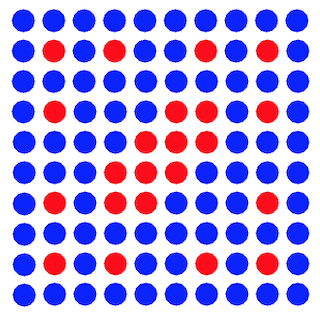}
\end{tabular}
\caption{Activity (left column) and attenuation (middle column) reconstructions and rod classification into present (blue) and missing (red) rods (right column) for the GE12 assembly \#12. The top row reconstructions are from data measured at normal height and the bottom row at the upper position, where the 14 partial fuel rods disappear from view.}
\label{fig:O25}
\end{figure}

A GE12 assembly contains two 2x2 water channels and 14 partial rods. Fig.~\ref{fig:O25} shows the attenuation and activity reconstructions and rod classification for the GE12 assembly \#12 (BU 43.1 GWd/tU, CT 11.7 a) both at the normal measurement height and in the upper position, where the partial rods disappear from view. Water channels are visible and correctly classified as missing, as are the partial rods from the higher measurement.

% O25 (#12)
% GE12 (BU 43.1 GWd/tU, CT 11.7 a)
% -1.5m as well

\begin{figure}[h!]
\centering
\begin{tabular}{@{}c@{\quad}c@{\quad}c@{}}
\includegraphics[scale=0.35]{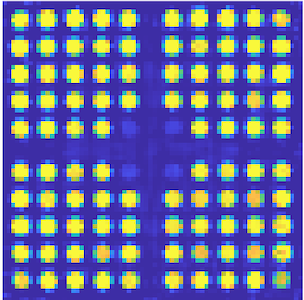}
&  \includegraphics[scale=0.35]{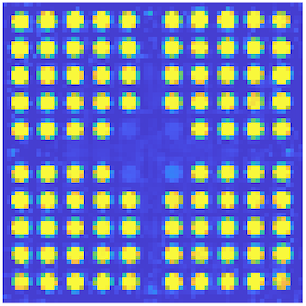}
& \includegraphics[scale=0.34]{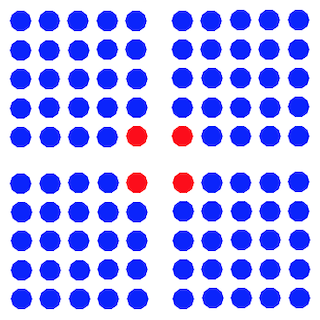}\\
\includegraphics[scale=0.35]{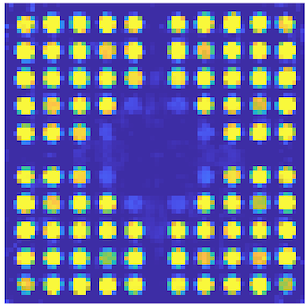} 
&  \includegraphics[scale=0.35]{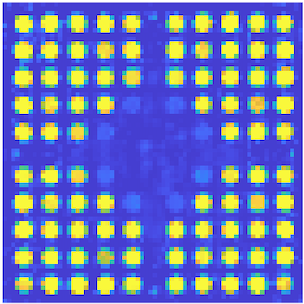}
& \includegraphics[scale=0.34]{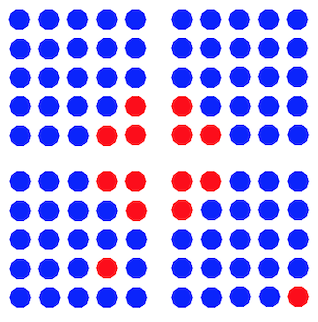}
\end{tabular}
\caption{Activity (left column) and attenuation (middle column) reconstructions and rod classification into present (blue) and missing (red) rods (right column) for the SVEA-96 OPTIMA assembly \#8. The top row reconstructions are from data measured at normal height and the bottom row at the upper position, where the partial fuel rods disappear from view. Note the two misclassified rods in the upper position measurements, see Section~\ref{sec:misclassifications}.}
\label{fig:O19}
\end{figure}

A SVEA-96 OPTIMA assembly contains 4 rod positions of water channel in the center and 8 partial rods next to them. Fig.~\ref{fig:O19} shows the SVEA-96 OPTIMA assembly \#8 (BU 39.8 GWd/tU, CT 13.7 a) at the normal measurement height as well as in the upper position where the partial rods disappear from view. The water channel positions in the center are correctly classified as missing as are the 8 partial rod positions in the upper position. Note that in the reconstructions from the upper position measurements there is one extra misclassified rod at the right lower corner and another in the lower-left part, see Section~\ref{sec:misclassifications} for details.

% O19 (#8)
% SVEA-96 OPTIMA (BU 39.8 GWd/tU, CT 13.7 a)
% also -1.5m, one "missing"

\subsection{Misclassified rods}
\label{sec:misclassifications}

On rare occasions, some rods are falsely classified as missing. The classification algorithm concludes that a certain fuel rod is missing based on limits for activity deviation from the rod's neighbors. The limits are defined by support vector machines trained with labelled training data (see Section~\ref{sec:analysis_reconstruction}).

In our studies, several different types of erroneous classifications occur. The results strongly suggest that one of the types of misclassifications is related to nearby water channels or partial rods which make it harder to detect whether a rod is present or missing. Another type is related to simplifications in the assembly geometry made before image reconstruction. Sometimes overall poor data quality, for example low gamma counts in certain measurements, cause activity and attenuation variation in the reconstruction and can lead to misclassifications.

An example of a faulty classification caused by the assembly geometry simplifications can be seen in Fig.~\ref{fig:O19}, where the reconstruction of the SVEA-96 OPTIMA fuel assembly \#8 is shown both at normal measurement height and in the upper position. The lower-right corner rod of the assembly is classified as missing in the case where the partial rods are not in view, although the rod is present. The reconstruction in the lower row is more uneven and thus prone to misclassifications.

\begin{figure}[h!]
\centering
\begin{tabular}{@{}c@{\quad}c@{\quad}c@{}}
\includegraphics[scale=0.5]{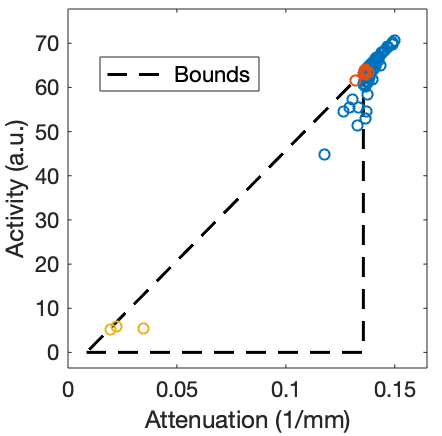} 
& \includegraphics[scale=0.5]{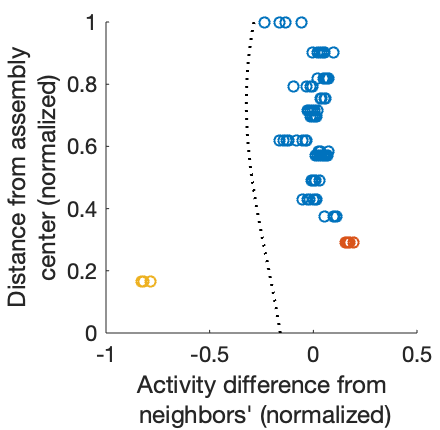} \\
\includegraphics[scale=0.5]{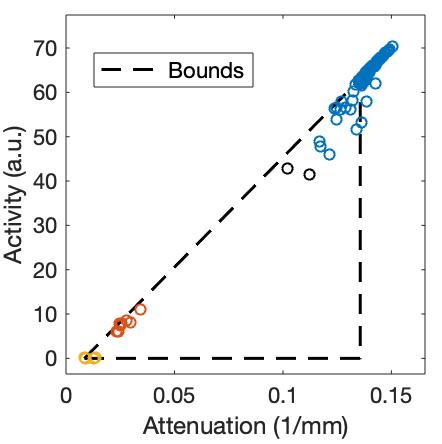} 
& \includegraphics[scale=0.5]{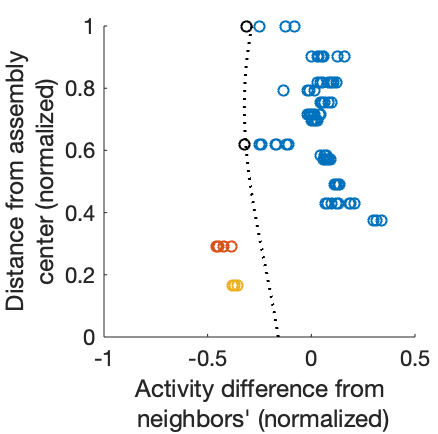}
\end{tabular}
     \caption{Rod classification plots for the SVEA-96 OPTIMA assembly \#8 (see also Fig.~\ref{fig:O19}), normal measurement height in the top row and upper position in the bottom row. Linear bounds and average rod values are shown in the attenuation-activity plane on  the left and rod activity difference from its neighbors as a function of the  distance from the assembly center is shown on the right. Circles represent individual rods and colors denote the ground truth rod type (blue for present, yellow for water channel, red for partial rod and black for rod that has been misclassified). The dotted line on the right represents the classification border.} 
     \label{fig:O19_bounds_neighbors}
\end{figure}

Fig.~\ref{fig:O19_bounds_neighbors} shows rod classification plots for the assembly \#8. In the same way as in Fig.~\ref{fig:12_bounds_neighbors}, rod positions with similar characteristics get grouped together. The water channels show low activity and attenuation and thus resemble missing rods. In the measurements collected at the upper position, the partial rods behave like water positions and end up with the water channel rods, and at normal measurement height they behave like present rods as they should. The black circles in the rightmost figure of the bottom row are the misclassified rods in the assembly (see Fig.~\ref{fig:O19}). As is clear, the margin to present rods is very small and a minor change in the classification border would result in a correct classification of the rods.

% O19, normal and 1.5m, one "missing" higher

The assembly geometry simplification causing corner rods getting misclassified can be observed in other reconstructions in addition to the previously presented Fig.~\ref{fig:O19}. The problem is limited to a certain assembly type (SVEA-types, especially SVEA-96) and is worst with the data gained from measurements at Olkiluoto 2017. The root cause for this is not yet completely confirmed, but it is related to the corner rods in these types of assemblies being smaller than the other rods and being placed closer to the center of the assembly in order to round the corners. This effect can be observed in the reconstructions.

\subsection{Intra-rod activity intensity deviations}
\label{sec:rod-level}

\begin{figure}[h!]
\centering
\begin{tabular}{@{}c@{\quad}c@{\quad}c@{}}
\includegraphics[scale=0.35]{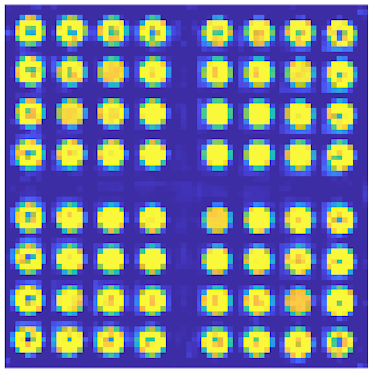}     
& \includegraphics[scale=0.35]{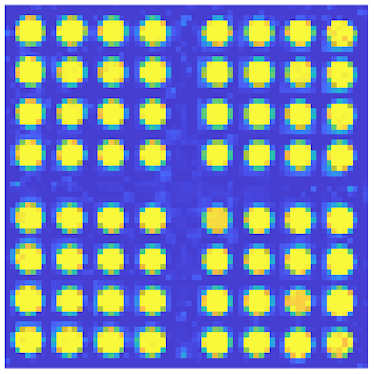}    
\end{tabular}
     \caption{Activity (left) and attenuation (right) reconstructions for a SVEA-64 assembly \#6, showing intra-rod activity intensity differences.} 
     \label{fig:O30}
\end{figure}

In some cases, especially with thicker rod diameters, the reconstructions show intra-rod activity differences. Fig.~\ref{fig:O30} shows the SVEA-64 assembly \#6 (BU 32.9 GWd/tU, CT 20.7 a), where these differences can be clearly seen as darker spots inside the outermost rods of the activity image.

The intra-rod activity differences are limited to SVEA-64 assemblies and a few 9x9-1AB and 8x8-1 assemblies. These assemblies all have in common a larger pellet diameter (9.5-10.4 mm) than the other assemblies (7.5-8.9 mm). A very small amount of the same phenomenon can be seen at the outer edge of some rods in the activity reconstructions in Fig.~\ref{fig:O1&O28} for the assembly \#5 and in Fig.~\ref{fig:O37&O41} for the assembly \#11.

A similar phenomenon has been studied by Caruso et al.~\cite{Caruso2009}. The intra-rod caesium and europium isotopic distributions were determined by gamma-activity tomography in high-burnup PWR fuel rods. Full-power irradiation at high temperatures causes fission products to diffuse from the hotter central region of the rod to the colder periphery, resulting in a non-uniform distribution of fission products inside the rod. Especially $^{134}$Cs and $^{137}$Cs diffuse easily and can cause significant differences in the fissile material content inside the rod. This phenomenon could explain our results, and the hypothesis is supported by the attenuation reconstruction, which does not show intra-rod differences. If the hypothesis holds, our results show promise in detecting even intra-rod distributions of nuclear material.

It should be noted that our chosen regularization terms (see Section \ref{sec:analysis_reconstruction}) are such that they prefer solutions where the intra-rod activity values and attenuation coefficients are uniform. This means that some of the intra-rod differences in the values are smoothed out in the reconstructions.

% O30 (#6)
% SVEA-64 (BU 32.9 GWd/tU, CT 20.7 a)
% dark spots inside rods

\subsection{Activity and attenuation gradient}
\label{sec:gradient}

Some assemblies show clear activity and attenuation gradients in the reconstructions. The gradual change of activity values throughout the assembly is normal to certain types of assemblies and is caused by their irradiation history and their placement in the reactor core during operation. The gradient on the attenuation coefficients, on the other hand, is an artefact of the reconstruction algorithm.

From the perspective of correct rod classification, the assemblies with a gradient demonstrate the ability of the algorithm to perform well even with more difficult assemblies. The gradient could cause some rods to be misclassified, because the algorithm evaluates rods by comparing their activity values to the values of their neighbors. Thus, significantly different values on opposite sides of the evaluated rod might cause the rod to be faultily classified. However, our results show that in most cases the smooth change in the activity throughout the fuel assembly does not cause misclassified rods.

\begin{figure}[h!]
\centering
\begin{tabular}{@{}c@{\quad}c@{\quad}c@{}}
\includegraphics[scale=0.35]{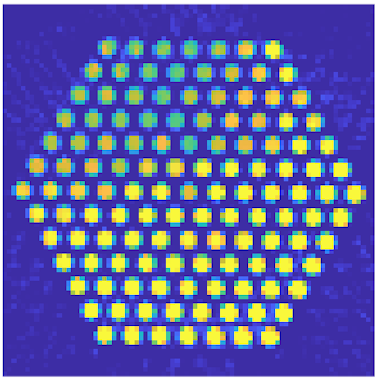}     
& \includegraphics[scale=0.35]{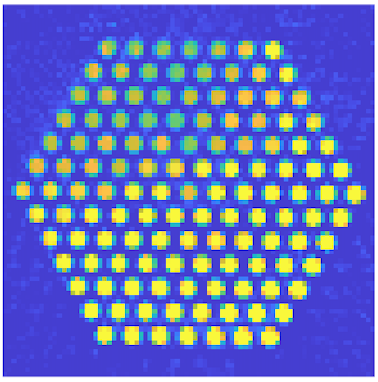}
& \includegraphics[scale=0.35]{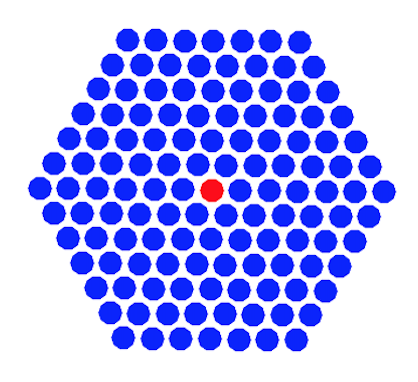}
\end{tabular}
     \caption{Activity (left) and attenuation (middle) reconstructions and classification into missing (red) and present (blue) rods (right) for the VVER-440 assembly \#4 with a water channel and a visible gradient in activity as well as attenuation. The water channel in the middle is correctly classified as missing.}
     \label{fig:28}
\end{figure}

An example of an assembly with a gradient can be seen Fig.~\ref{fig:28}, where the activity and attenuation reconstruction and rod classification of the VVER-440 assembly \#4 (BU 22.8 GWd/tU, CT 27.6 a) is shown.

% 28 (#4)
% VVER-440 (BU 22.8 GWd/tU, CT 27.6 a)
% gradient, center water channel noticed

\section{Discussion}

The results presented in Section~\ref{sec:results} show that the reconstruction method gives a reliable basis for the classification of fuel rods. However, our current method of classifying the rods in only two distinct categories (missing and present) is somewhat insufficient for the safeguard purposes at the disposal in a geological repository. There are more possibilities for nuclear material diversion than just removing a whole rod. The rod can be for example partially removed or replaced with another material. Related to this, also the burnable absorber rods should be somehow notified by the classification because they behave abnormally as well. The burnable absorber rods show that it is actually very difficult to detect rods that are replaced by a material with similar characteristics to spent fuel. Especially with low burnup it is almost impossible to detect these kind of deviations.

In the case of burnable absorber rods, we know from the licence-holder's declaration that these rods might have a lower burnup and we can respond to the classification result accordingly. However, in other cases such an abnormal result might indicate a replacement of the rod with inactive material of the same density as the fuel or a diversion of a part of the fuel rod. Further investigation could then be done. Replacement scenarios and how to detect them are a topic of future research.

For the above-mentioned reasons there is a need to create additional classification categories to account for cases where the fuel rod might be modified or replaced with some other material. Division could be done into present, abnormal and missing rods and the categories could be given a different priority depending on actions needed. Abnormal assemblies would require further investigations to get the "green light" to proceed with the disposal.

An example of the differing behaviour indicating something abnormal is given by assembly \#5 in Fig.~\ref{fig:O1&O28}, where the abnormal fuel rods noticed by the algorithm are present but have a lower burnup than the other rods in the assembly. In the current version of the software, these rods are classified as missing due to their low activity value. Still, the presence of the rods in the assembly is clear from the attenuation image, although the attenuation coefficients also differ slightly from the other rods. In the future version of the classification algorithm, the rods could be classified as "modified" and then investigated in more detail. This would also reduce the possibility of false alarms.

There are some aspects of the software that need revision. The misclassifications caused by geometry assumptions, as presented in Section~\ref{sec:misclassifications}, will be addressed to ensure that no false alarms will be given due to simplifications in the geometry assumptions. The issue is demonstrated by the reconstructed activity and attenuation plots in Fig. ~\ref{fig:O19}. The corner rods of the assembly can be observed to be somewhat smaller than the other nearby rods. The placement of the corner rod is also a bit different from the other rods in the same row and column such that the overall shape of the assembly is a bit rounded. The reason behind this geometrical arrangement is to even out the reactivity and neutron fluxes in the reactor during operation. The geometry assumptions of this assembly type will be revised and implemented in the software.

We see room for improvement in the determination of the pre-defined activity and attenuation bounds. The presently used triangular bounds, e.g., do not contain high-activity surrogate rods made from less attenuating material such as steel. However, opening up the bounds too much leads to poorer images; an optimum solution will thus need to be found.
We are working on making the forward model more realistic by adding gamma ray scattering. Presently, only absorption is taken into account.
In the context of incorporating PGET in the geological disposal safeguards activity, the processes of data acquisition, image reconstruction and rod classification will be integrated and automated.

\section{Conclusion}

The presented results show that the simultaneous reconstruction of activity and attenuation images works as a reliable base for fuel rod classification. The developed method is confirmed with data from a wide range of spent fuel assembly types and parameters measured at the Finnish nuclear power plants. The method is well suited for nuclear safeguards verification of fuel assemblies in Finland prior to deep geological disposal.

We are working on improving the reconstruction method and the classification algorithm. The classification criteria will be revised to include a further category for rods that might be modified but not missing. Other future work includes improving the activity-attenuation bound estimation method and including scattering in the forward model to describe the physical phenomena inside the fuel assembly more realistically.

\section{Acknowledgements}

TAB acknowledges support by the Academy of Finland postdoctoral grant, decision number 330522.
TAB and SS acknowledge partial support by the Academy of Finland through the Finnish Centre of Excellence in Inverse Modelling and Imaging 2018-2025, decision number 312339.
PD acknowledges partial support by Business Finland under Grant 1845/31/2014.
TH was supported by the Academy of Finland via the decision number 326961.

\bibliographystyle{IEEEtran}
\bibliography{Bibliography}

\end{document}